\documentclass{article}
\usepackage{spconf,amsmath,graphicx}
\usepackage[colorinlistoftodos]{todonotes}
\usepackage{caption}
\usepackage{subcaption}
\usepackage{bm}
\usepackage{hyperref}
\usepackage{xcolor}
\usepackage{color}
\usepackage{bm}
\usepackage{url}
\usepackage{subcaption}
\usepackage{booktabs}
\usepackage{multirow}
\usepackage{textcomp}
\usepackage{verbatim}
\usepackage{amssymb}
\hypersetup{
    colorlinks=true,
    linkcolor=blue,
    filecolor=magenta,      
    urlcolor=blue,
}

\usepackage{soul}
\usepackage{color}
\usepackage{xcolor}

\newcommand\omitthis[1]{}

\title{Binaural Angular Separation Network}

\name{Yang Yang, George Sung, Shao-Fu Shih, Hakan Erdogan, Chehung Lee, Matthias Grundmann}

\address{Google LLC, U.S.A.}
\begin{document}
%
\maketitle
\vspace{-1cm}
\begin{abstract}
We propose a neural network model that can separate target speech sources from interfering sources at different angular regions using two microphones. The model is trained with simulated room impulse responses (RIRs) using omni-directional microphones without needing to collect real RIRs. By relying on specific angular regions and multiple room simulations, the model utilizes consistent time difference of arrival (TDOA) cues, or what we call delay contrast, to separate target and interference sources while remaining robust in various reverberation environments. We demonstrate the model is not only generalizable to a commercially available device with a slightly different microphone geometry, but also outperforms our previous work which uses one additional microphone on the same device. The model runs in real-time on-device and is suitable for low-latency streaming applications such as telephony and video conferencing.
\end{abstract}

\vspace{-0.6cm}
\section{Introduction}
\label{sec:intro}
\vspace{-0.2cm}

Audio source separation is used in many applications from voice communication to human computer interface. Previous works on multi-channel speech separation and enhancement focused on using spatial and spectral cues to separate sources arbitrarily mixed with each other. Works such as deep beamforming networks \cite{xiao_2016,li_icassp2022}, neural beamforming \cite{heymann2016neural, erdogan2016improved, wang2021sequential, xu2021generalized, wang2023tf}, or other direct multi-channel separation methods \cite{luo2019fasnet,yoshioka2022vararray} focus on a more general separation problem and do not typically assume an information or assumption about locations of sources.

To achieve audio separation, one method is to utilize spatial cues with multiple microphones using what is known as beamforming. Due to the nature of linear processing, conventional beamforming performance highly depends on the number of microphones and is limited in terms of suppression of interference and enhancement of target signals \cite{benesty2008microphone}. In addition, it is not possible to control the angle ranges easily in beamforming since main-lobe width and side-lobe levels are typically variable for each frequency. Beamformer modules have been used in neural beamforming \cite{wang2021sequential, xu2021generalized, wang2023tf} to linearly re-estimate initially separated targets using multi-microphone data or in GSENet \cite{gsenet} to provide magnitude contrast to a neural network for further refined separation of the target source within a mixture.

There have been works relying on explicitly using locations of sources, such as location-guided separation \cite{location_guided} which 
aims to separate a source in any possible angle from a mixture with a known microphone geometry, specifically using a circular array. Another related work is distance-based separation \cite{distance_based_separation} which is designed to separate sources within a designated distance from one single microphone. The network used a single microphone and relied on impulse response characteristics for separation rather than using inter-microphone cues. In recent works, angular positions of sources are used to order individual speech sources \cite{location_based_ordering} to avoid Permutation-Invariant Training (PIT) for speech separation. Neural Spectro-Spatial Filtering\cite{tan2022neural} performs separation of target and interference signals either through location-based ordering similar to \cite{location_based_ordering} or by assuming a singular location for a target speech source with multiple possible locations for a noise source. {\color{black} In \cite{wechsler2023icassp}, the authors propose a region-based separation method to separate in-car audio into rectangular regions using a 3-mic linear array. Due to region shapes, sources would not cause consistent TDOA cues which could make it harder for the network to separate regional signals.}

In this paper, we propose a new end-to-end paradigm including simulator design, model training, and on-device inference for two microphones {\color{black} angular region based} source separation applications. {\color{black} We name our model as {\bf BASNet}, short for {\bf B}inaural {\bf A}ngular {\bf S}eparation {\bf Net}work\footnote{Audio samples are available at \href{https://google-research.github.io/seanet/basnet/}{google-research.github.io/seanet/basnet/}}.} The model assumes the target and interference sources are located within specific angle ranges. This assumption allows the network to implicitly focus on inter-microphone phase differences (IPD) or time difference of arrival (TDOA) information to separate the sources in a reliable manner.  TDOA cues remain consistent throughout training due to using fixed target and interference angle ranges and it makes it easy for the network to rely on that information to perform separation. We train the network using room simulations based on the image method. The benefit compared to other methods come from the network's capability to seamlessly combine spectro-spatial information over all microphones and frequencies, as well as being robust to the reverberation environment through extensive simulated training. The trained network generalizes well to real-world data recorded in a lab. Additionally, the training process does not require on-device data collection, which is another advantage over previous methods such as GSENet \cite{gsenet}. We show that our method achieves better separation performance than previous neural beamforming methods. In particular, using two microphones with our method provides a significant performance gain over using a single microphone, unlike previous methods such as Sequential Neural Beamforming \cite{wang2021sequential}. In contrast to the Location-Based Training\cite{location_based_ordering} which orders output speakers according to their angles, our proposal is based on a fixed range of angles for target and interference and aim to separate target speech from speech and non-speech interference. Unlike \cite{tan2022neural} which used simulations with measured or simulated RIRs for evaluation, we evaluate our method on real recorded examples.

\begin{figure}[!t]
\centering
\includegraphics[width=0.35\textwidth]{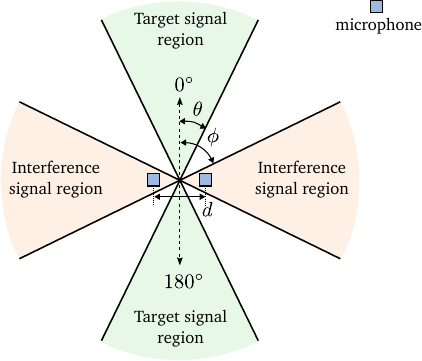}
\caption{\small RIR simulation setup for target and interference sources. Target signal sources is confined to $[-\theta, +\theta]$ and $[180^{\circ}-\theta, 180^{\circ}+\theta]$; interference sources is confined to $[90^{\circ}-\phi, 180^{\circ}-\phi]$ and $[180^{\circ}+\phi, 360^{\circ}-\phi]$. Noise source can come from any of the $360^{\circ}$ directions. Distance between the two microphones is denoted as $d$.}
\label{fig:rir}
\end{figure}

\begin{table}[t]
\setlength{\tabcolsep}{7pt}
\footnotesize
\caption{Data pipeline parameter setup.}
\label{tab:config}
\centering
\begin{tabular}{  l l  } 
\toprule
Type  & Configuration \\
\midrule
Geometry& $\theta=30^{\circ}$, $\phi=60^{\circ}$, $d\sim \text{Uniform}[0.09m, 0.11m]$ \\
\hline
\multirow{3}{4em}{Signal synthesis}& $p_1=0.8$, $p_2=0.6$ \\
& $g_0 \sim \mathcal{N}(0, 0)$, $g_1\sim \mathcal{N}(-3, 3)$, $g_2\sim\mathcal{N}(-3, 3)$, \\
& $g_3\sim\mathcal{N}(-5, 10)$,  $g_{\text{global}}\sim\mathcal{N}(-10, 5)$ \\
\bottomrule
\end{tabular}
\vspace{-0.4cm}
\end{table}

\vspace{-0.5cm}
\section{Method}
\label{sec:method}
\vspace{-0.3cm}

\subsection{Data Pipeline and Training}
\label{ssec:datapipeline}
\vspace{-0.2cm}

To utilize the spatial cues for the model by contrasting two audio inputs, we design an input simulator which generates room impulse responses (RIR) to synthesize two input channels.

The simulator generates rooms with different geometries. For each room, two microphone locations with distance $d$ are randomly sampled, where $d$ follows a predefined distribution. Based on the location of the microphones, the space is divided into three disjoint regions as illustrated in Fig.~\ref{fig:rir}: the target signal region consists of points in the 3-dimensional space where the angle of its position vector\footnote{Origin of 3D Cartesian space is defined as the midpoint between mics.} to $0^{\circ}$ plane relative to the two microphones is less than $\theta$. The $0^{\circ}$ plane goes through the mid point of the two microphones and is orthogonal to the line connecting the two. The interference signal region contains points whose position vectors are at least $\phi$ degree from the $0^{\circ}$ plane. Four signal sources are created where two target speech sources are randomly sampled in the target signal region, one interference speech source is sampled in the interference signal region, and a noise source is sampled randomly unconstrained by the two regions. {\color{black} Delay contrast occurs because direct path far-field response of the sources in the target and interference regions consistently achieve a distinct range of TDOAs (or relative delays) and the network can rely on these cues for separation.}

With the room geometry, 2 microphones locations, and 4 signal source locations determined, a $4\times2$ RIR matrix\footnotemark $\{\bm{r}_{(k, j)}\}_{0\leq k \leq 3, 0\leq j \leq 1}$ is created using the image method \cite{allen_image_method_1979}. The raw audio capture from the two microphones are synthesized following the equation below\footnote{\color{black}$*$ denotes convolution.}:
\begin{align}
    \bm{y}_0 = &  \bm{s}_1 * \bm{r}_{(0,0)} + \bm{s}_2 * \bm{r}_{(1,0)}  +  \bm{i} * \bm{r}_{(2,0)} + g_n \cdot\bm{n} * \bm{r}_{(3,0)},  \notag\\
    \bm{y}_1 = & \bm{s}_1 * \bm{r}_{(0,1)} + \bm{s}_2 * \bm{r}_{(1,1)}  +  \bm{i} * \bm{r}_{(2,1)} +  g_n \cdot\bm{n} * \bm{r}_{(3,1)}, \notag
\end{align}
where $\bm{s}_1$, $\bm{s}_2$ and $\bm{i}$ are utterances from a speech dataset, and $\bm{n}$ comes from a noise dataset. With probability $p_1$, the utterance $\bm{s}_2$ is set to empty. With probability $p_2$, the utterance $\bm{i}$ is set to empty. The introduction of $p_1$ and $p_2$ is to ensure the model can handle both single and multiple target speeches as separation target, and both with and without the presence of interference speech. To add variations to the signal strengths of different components, the average power of the four components are controlled by normalizing and scaling the signal to follow a sampled dB value, denoted as $\{g_k\}_{0\leq k\leq 3}$. A global power normalizing and scaling is then applied to set the final output power to be $g_{\text{global}}$. The exact numerical configurations for the data pipeline is reported in Table~\ref{tab:config}. The ground-truth signal for model training is the non-reverberated version of the input without the presence of noise and interference sources, derived following the equation below\footnote{\color{black}$\text{anechoic}(\cdot)$ denotes the anechoic version of the RIR, which only contains the strongest path.}.
\begin{align}
    \bm{t} = & \bm{s}_1 * \text{anechoic}(\bm{r}_{(0, 0)}) + \bm{s}_2 * \text{anechoic}(\bm{r}_{(1, 0)}). \notag
\end{align}

\footnotetext{The RIRs are normalized so that the magnitude of the largest peak, over all receivers, for each source, is 1.}

\vspace{-0.5cm}
\subsection{Model Architecture}
\label{ssec:modelarch}
\vspace{-0.2cm}

The model utilizes a convolution U-Net with identical architecture with GSENet (see Fig.~2 in \cite{gsenet} for details). The input to the network are the STFTs of the two raw microphone inputs packed in real and imaginary channels, and output is the STFT of the reconstructed waveform followed by an inverse-STFT to convert to waveform. The input STFT and output inverse-STFT have a window size (the same as fft size) of 320 and step size of 160.  A single-scale STFT re-construction loss \cite{engel_ddsp_iclr_2020} with window size of 1024 and step size of 256 is applied on the reconstructed waveform. 

The network is fully causal with the coarsest temporal resolution in the U-Net be limited at 2 times that of the input. At inference time, the network can be applied in a streaming fashion with a latency of 20ms (320 samples at 16kHz) using the streamable library in \cite{kws_streamable}. 

\vspace{-0.4cm}
\subsection{Real Time Inference}
\vspace{-0.3cm}
\label{ssec:inference}

Since the network architecture is identical to GSENet \cite{gsenet}, the proposed method inherits the same real-time inference capability (for details see Section~2.3 in \cite{gsenet}), with an average latency of 31.81ms profiling on a single CPU core of a Pixel 6 model phone with the XNNPACK backend \cite{xnnpack}.

\vspace{-0.4cm}
\section{Experiments}
\label{sec:pagestyle}
\vspace{-0.3cm}

\subsection{Training and Evaluation Dataset}
\vspace{-0.2cm}
Both target speech {\color{black}$\bm{s}_1, \bm{s}_2$} and interference speech $\bm{i}$ are sampled from a combination of LibriVox \cite{librivox} and internal speech datasets. Background noise $\bm{n}$ is sampled from Freesound \cite{freesound} dataset. For all the experiments, all the files are resampled to 16kHz sampling rate.

\begin{figure}[!t]
\centering
\includegraphics[width=0.35\textwidth]{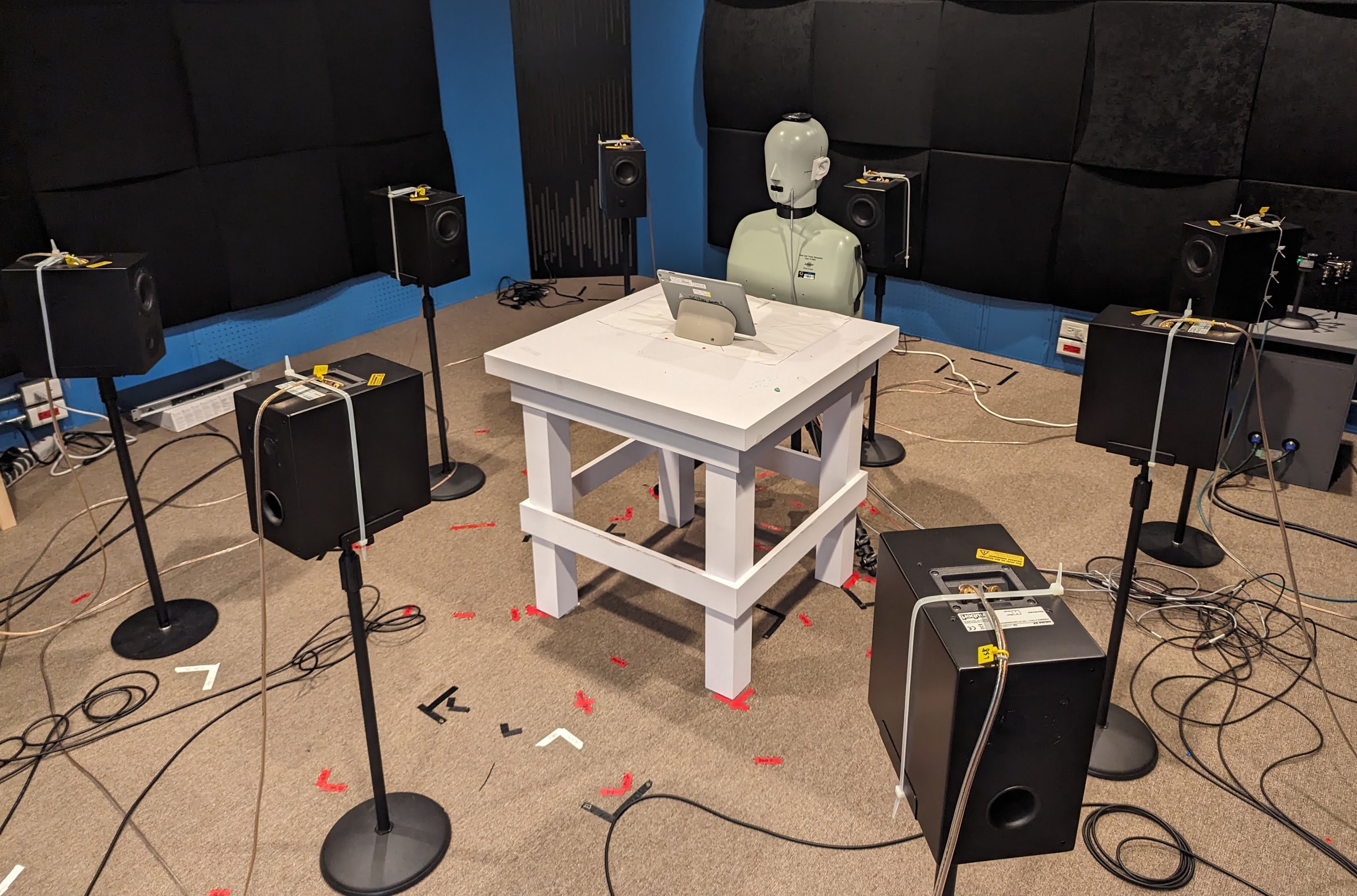}
\caption{Listening room setup.}
\vspace{-0.3cm}
\label{fig:lab}
\end{figure}

For evaluation, we use multi-channel audio collected from a Google Pixel Tablet \cite{pixel_tablet} in an ETSI certified listening room as shown in Fig.~\ref{fig:lab}. The tablet is docked on the speaker dock and secured on the table, referred as device-under-test (DUT). The DUT is equipped with 3 microphones: two of them on the top edge with 0.07 meter symmetrical spacing to the center, and one on the right edge. A head-and-torso simulator (HATS) is placed at 0 degree in front of the DUT to simulate the target speech source. For directional evaluation, 8 individual loudspeakers are placed surrounding the device with $45^{\circ}$ succession which represent the ambient noises or interference speech sources. Note that the $0^{\circ}$ speaker is placed behind the HATS. For each loudspeaker, DEMAND noise \cite{thiemann_demand_dataset_2013} data and a subset of VCTK speech \cite{yamagishi_vctk_2019} data are played and recorded from the DUT. Each of the speaker and HATS recordings are done independently and later mixed with various mixtures and SNRs conditions for model evaluation. To measure the directivity pattern of the processed results, we record another set of audio without the HATS hence the $0^{\circ}$ speaker is unobstructed.

\vspace{-0.4cm}
\subsection{Evaluation Methods}
\vspace{-0.3cm}

The comparison baselines use all 3 microphones. {\bf MCWF} is a DSP-based beamformer based on linear multi-channel Wiener filter introduced in \cite{wang2021sequential}, where a small amount of recorded data is used to derive the beamformer weights. More specifically, the signal covariance matrix is derived from the HATS recording while the noise covariance matrix is derived from the loudspeaker recordings excluding the $0^{\circ}$ and $180^{\circ}$ directions (refer to Section 3.2 in \cite{gsenet} for details). Additionally, we compare to two other baselines: {\bf SSENet} \cite{gsenet}, a single channel speech enhancement network after the beamformer; and {\bf GSENet} \cite{gsenet}, a speech enhancement network that takes both beamformer output and a stream of raw microphone as input.

GSENet leverages the \textbf{\emph{magnitude contrast}} between two inputs: the 3-channel beamformed output as the target speech input, and  one of the raw microphones as the comparison input. GSENet operates on the assumption that, based on MCWF assumption, the target beamformed speech from $0^{\circ}$ is distortion-less while signals from other interference directions are attenuated. Whereas BASNet uses only the top two microphones with symmetrical placement as raw inputs, and mostly relies on \textbf{\emph{delay contrast}} {\color{black}-- the consistent time difference of arrival (TDOA) cues --} to separate speech signal at target angle from interference angles.
\vspace{-0.45cm}
\subsection{Evaluation Results}
\vspace{-0.3cm}

\begin{table*}[t]
\setlength{\tabcolsep}{3.6pt}
\small
\caption{
 BSS-SDR (dB) \cite{raffel_bss_sdr_2014} of the raw and enhanced speech waveform with interference coming from different angles. Left: speech as interference. Right: noise as interference. Top: interference at 0dB SNR. Bottom: interference at 6dB SNR.
}
\vspace{-0.1cm}
\label{table:bss_sdr}\vspace{-0.2cm}
\centering
\begin{tabular}{ @{}l c c c c c c c c c c c c c c c c c c c @{}} 
\toprule
& \multicolumn{9}{c}{Speech (VCTK) as interference} & \multicolumn{9}{c}{Noise (DEMAND) as interference}\\
 \cmidrule(lr){2-10} \cmidrule(lr){11-19}
Interference angle   & 0\textdegree & 45\textdegree & 90\textdegree & 135\textdegree & 180\textdegree & 225\textdegree & 270\textdegree & 315\textdegree & avg.  & 0\textdegree & 45\textdegree & 90\textdegree & 135\textdegree & 180\textdegree & 225\textdegree & 270\textdegree & 315\textdegree & avg.\\
\midrule
BF (MCWF) \cite{wang2021sequential} &0.5 & 2.3 & 3.0 & 1.8 & 0.2 & 1.7 & 2.2 & 1.9 & 1.7 & 4.1 & 6.3 & 5.9 & 4.4 & 3.2 & 4.8 & 5.7 & 5.0 & 4.9 \\
BF + SSENet \cite{gsenet} & 0.6 & 2.4 & 3.1 & 1.9 & 0.2 & 1.9 & 2.5 & 2.1 & 1.8 & 10.3 & 12.6 & 12.1 & 10.7 & 9.6 & 11.1 & 12.0 & 11.3 & 11.2 \\
BF + GSENet \cite{gsenet} & 0.8 & 7.7 & 10.5 & 9.0 & 0.2 & 6.1 & 9.8 & 9.1 & 6.7 & 9.6 & 13.0 & 12.7 & 11.3 & 8.8 & 11.8 & 12.6 & 11.8 & 11.4\\
BASNet (ours) & 3.9 & 11.3 & 13.1 & 11.7 & -0.1 & 8.5 & 12.4 & 12.0 & 9.1 & 11.6 & 14.8 & 14.0 & 12.7 & 10.7 & 12.5 & 13.9 & 13.3 & 12.9\\
\midrule
BF (MCWF) \cite{wang2021sequential}  & 6.3 & 7.9 & 8.6 & 7.5 & 6.0 & 7.4 & 7.9 & 7.5 & 7.4 & 9.5 & 11.4 & 11.1 & 9.8 & 8.7 & 10.1 & 10.9 & 10.2 & 10.2\\
BF + SSENet \cite{gsenet}& 6.3 & 8.0 & 8.6 & 7.5 & 5.9 & 7.5 & 8.0 & 7.6 & 7.4 & 13.4 & 14.7 & 14.5 & 13.6 & 12.8 & 13.8 & 14.4 & 14.0 & 13.9\\
BF + GSENet \cite{gsenet}  & 6.4 & 11.4 & 13.1 & 12.0 & 5.9 & 10.3 & 12.6 & 12.0 & 10.5 & 12.6 & 14.9 & 14.8 & 13.9 & 11.7 & 14.2 & 14.7 & 14.2 & 13.9\\
BASNet (Ours) & 8.2 & 15.2 & 16.5 & 15.3 & 5.9 & 12.3 & 15.7 & 15.3 & 13.1 & 15.5 & 18.0 & 17.4 & 16.2 & 14.8 & 16.2 & 17.2 & 16.8 & 16.5\\
\bottomrule
\end{tabular}
\vspace{-0.3cm}
\end{table*}

The evaluations are done with two criteria: enhancement and steerability. To evaluate speech enhancement effectiveness, the setup is configured to have target signal present from HATS at $0^{\circ}$ consistently, while the interference is arbitrarily assigned to a combination of 8 loudspeakers. For steerability, we use the recording collected without HATS to evaluate the interference only performance, and showcase the model's capability to steer to different spatial directions by introducing artificial latency to one of its inputs.

\begin{figure}[!t]
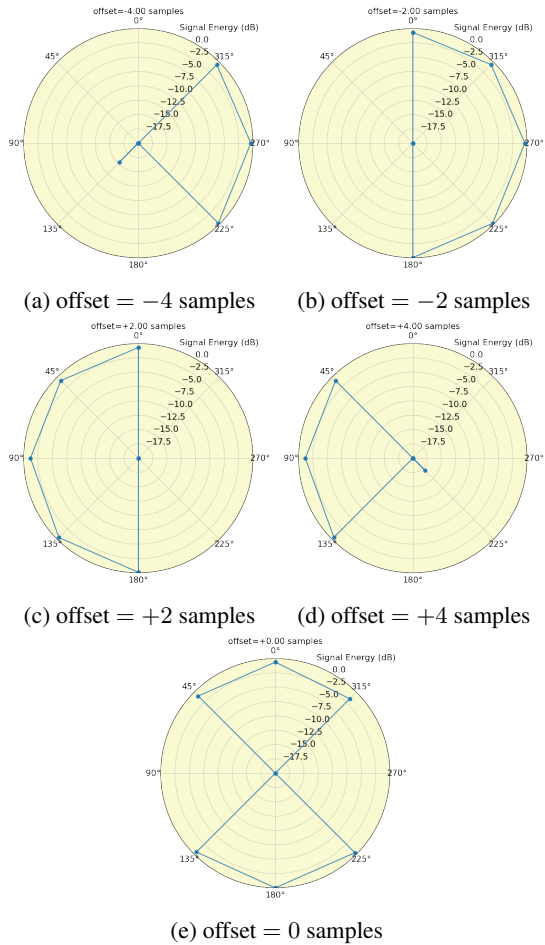

 \centering
 \begin{subfigure}[b]{0.20\textwidth}
     \centering
     \includegraphics[width=\textwidth]{figures/steering/00.pdf}
     \caption{offset $= -4$ samples}
     \label{fig:270degree}
 \end{subfigure}
 \begin{subfigure}[b]{0.20\textwidth}
     \centering
     \includegraphics[width=\textwidth]{figures/steering/10.pdf}
     \caption{offset $= -2$ samples}
     \label{fig:270wdegree}
 \end{subfigure}\\
   \begin{subfigure}[b]{0.20\textwidth}
     \centering
     \includegraphics[width=\textwidth]{figures/steering/30.pdf}
     \caption{offset $= +2$ samples}
     \label{fig:90degree}
 \end{subfigure}
 \begin{subfigure}[b]{0.20\textwidth}
     \centering
     \includegraphics[width=\textwidth]{figures/steering/40.pdf}
     \caption{offset $= +4$ samples}
     \label{fig:90wdegree}
 \end{subfigure}
 \begin{subfigure}[b]{0.20\textwidth}
     \centering
     \includegraphics[width=\textwidth]{figures/steering/20.pdf}
     \caption{offset $= 0$ samples}
     \label{fig:0degree}
 \end{subfigure}
 
\caption{Directivity pattern with different sample offsets.}
\vspace{-0.5cm}
\label{fig:directivity}
\end{figure}
\vspace{-0.4cm}
\subsubsection{Speech Enhancement with Directional Inference}
\vspace{-0.2cm}
In Table~\ref{table:bss_sdr}, we evaluate the scenario when there is target speech played from the HATS, and report the BSS-SDR \cite{raffel_bss_sdr_2014} of the model output while interference is played from each of the 8 loudspeakers. 

When the interference is speech, beamformer (BF) + SSENet performs similarly to the BF alone. Given that SSENet only has access to mono audio channel, this is the expected behavior as SSENet is not capable of separating the target speech from the interference speech based on locations. In contrast, BF + GSENet delivers an additional 3dB gain on average over BF + SSENet. To our surprise, BASNet, with only two microphone inputs, delivers another 2.4dB gain over BF + GSENet which utilizes three microphones.

When the interference is noise, BF + SSENet and BF + GSENet achieves similar performance, and the BSS-SDR values are close to uniform across all noise directions. In contrast, BASNet out-performs both by 1.5dB at 0dB SNR and 2.6dB at 6dB SNR, and generally performs better at $90^{\circ}$ and $270^{\circ}$ directions where the delay contrasts from the two microphones are the largest. This demonstrate the model's capability in utilizing delay (or TDOA) information to achieve better denoising performance.

\begin{table}[h]
\setlength{\tabcolsep}{3.6pt}
\footnotesize
\caption{
 Signal energy suppression (dB) when there is only one speech source coming from different angles.
}
\label{table:interference_only}\vspace{-0.2cm}
\centering
\begin{tabular}{ @{}l c c c c c c c c c @{}} 
\toprule
Angle   & 0\textdegree & 45\textdegree & 90\textdegree & 135\textdegree & 180\textdegree & 225\textdegree & 270\textdegree & 315\textdegree \\
\midrule
BF (MCWF) \cite{wang2021sequential} & 1.4 & 3.3 & 4.1 & 2.6 & 2.0 & 2.8 & 2.6 & 2.0 \\
BF + SSENet \cite{gsenet} & 1.6 & 3.7 & 4.4 & 2.9 & 2.1 & 3.1 & 2.9 & 2.3 \\
BF + GSENet \cite{gsenet}  & 1.6 & 10.0 & 18.2 & 7.3 & 2.2 & 16.9 & 16.0 & 3.0 \\
BASNet (ours)  & 0.6 & 1.0 & 46.4 & 0.5 & 0.0 & 0.4 & 44.1 & 1.6 \\
\bottomrule
\end{tabular}
\vspace{-0.4cm}
\end{table}

\vspace{-0.2cm}
\subsubsection{Steerable Directivity}
\vspace{-0.2cm}

In Table~\ref{table:interference_only}, we report the reduction of signal energy with only interference signals from each one of the 8 speakers without the presence of HATS, to measure the directivity pattern of the model. We observe that, compared to BF + GSENet which achieves a wider rejection region (rejection happens not just on $90^{\circ}$ and $270^{\circ}$), BASNet achieves much stronger $>40dB$ rejection at $90^{\circ}$ and $270^{\circ}$.

We postulate that BASNet preserves signals for which there is small relative delay in the two inputs, therefore we should be able to steer the direction of its directivity pattern by introducing artificial delay to one of its inputs. We verify this hypothesis as shown in Fig.~\ref{fig:directivity} that, by introducing sample offsets, BASNet can separate speech components from different directions. The ability to steer the focus of the model to different spatial regions allows the model to be dynamically adapted to target speakers using visual cues or manual inputs.

\vspace{-0.3cm}
\section{Conclusion}
\vspace{-0.3cm}
\label{sec:conclusion}

In this work, we propose a model that takes two audio channels as input and rely on the delay contrast between the two to preserve target speech and suppress interference ones. We show that, on a real device, it achieves state-of-the-art speech enhancement in the case of directional interference. We further demonstrate the steerability of its directivity pattern, which allows the same model to be used to adapt to different target spatial regions. For future work, we plan to explore how to utilize more than 2 microphone inputs, and how to combine magnitude contrast \cite{gsenet} and delay contrast to achieve even stronger enhancement and separation performance.
 


\let\oldbibliography\thebibliography
\renewcommand{\thebibliography}[1]{\oldbibliography{#1}
\setlength{\itemsep}{0pt}} 
\bibliographystyle{IEEEbib}
\bibliography{strings,refs}

\end{document}